\documentclass{aa}
\usepackage{graphicx}
\usepackage{txfonts}
\usepackage{natbib}
\begin{document}

\title{VLBA SiO maser observations of the OH/IR star OH~44.8-2.3: magnetic field and morphology}

\author{N. Amiri  \inst{1,3}
       \and
       W. H. T. Vlemmings
       \inst{2}
      \and
      A. J. Kemball  \inst{4}
       \and
        H.J. van Langevelde  \inst{3,1}
        }
  \institute{Sterrewacht Leiden, Leiden University, Niels Bohrweg 2, 2333 CA Leiden, The Netherlands
            \and
            Argelander Institute for Astronomy, University of Bonn, Auf dem H\"{u}gel 71, 53121 Bonn, Germany
            \and
            Joint Institute for VLBI in Europe (JIVE), Postbus 2, 7990 AA Dwingeloo, The Netherlands
            \and
            Department of Astronomy and Institute for Advanced Computing Applications and Technologies/ NCSA, University of Illinois at Urbana-Champaign,
1002 W. Green Street, Urbana, IL 61801, USA}
\date{Received  ; Accepted }

\abstract 
{SiO maser emission occurs in the extended atmosphere of evolved stars and can be studied at high angular resolution. As compact, high brightness components they can be used as important tracers of the dynamics at distances close to the central star. The masers also serve as probes of the evolutionary path from spherically symmetric AGB stars to aspherical PNe. Very long baseline interferometry (VLBI) observations of Mira variables indicate that SiO masers are significantly linearly polarized with linear polarization fraction up to 100$\%$. However, no information is available at high angular resolution for SiO masers in higher mass loss OH/IR stars. Theory indicates a different SiO pumping mechanism in higher mass loss evolved stars.}{We extend the VLBI SiO maser studies to OH/IR stars. The observations enable us to understand the SiO pumping mechanisms in higher mass loss evolved objects and compare those with Mira variables. Additionally, polarimetric observations of SiO masers help us to understand the magnetic field strength and morphology and to distinguish between conflicting polarization theories.} {The 43 GHz SiO maser observations of the OH/IR star OH~44.8-2.3 were performed with the VLBA in full polarization spectral line mode. Auxiliary EVLA observations were performed to allow for the absolute calibration of the polarization angle. The Zeeman splitting was measured by cross correlating the right and left circular polarization spectra as well as the S-curve fitting. Additionally, we analyzed the 1612 MHz OH maser observations of OH~44.8-2.3 from the VLA archive.}{The SiO masers of OH 44.8-2.2 form a ring located at $\sim$5.4 AU around the star. The masers appear to be highly linearly polarized with fractional linear polarization up to 100$\%$. The linear polarization vectors are consistent with a dipole field morphology in this star. We report a tentative detection of circular polarization of $\sim$0.7$\%$ for the brightest maser feature. The magnetic field measured for this feature corresponds to 1.5$\pm$0.3 G. Additionally, the distribution of the 1612 MHz OH maser emission could indicate an elongated morphology.} {}  
\maketitle
\maketitle

\section{Introduction}\label{introduction ch6}

SiO maser emission associated with Asymptotic Giant Branch (AGB) stars occurs at the inner region of Circumstellar Envelopes (CSEs). Maser emission has been detected in several vibrationally-excited rotational transitions of the SiO molecule \citep[e.g.][]{humphreys1997,pardo1998,soria2007}. A series of very long baseline array (VLBA) high resolution observations of the SiO masers around Mira variables have shown that the maser emission is confined to a region, sometimes ring-shaped, between the stellar photosphere and the dust formation zone \citep[e.g.][]{cotton2008, cotton2006, diamond1994}. On scales probed by very long baseline interferometry (VLBI), the masers are confined to localized spots with lifetime of a few months. The resolution of the VLBI observations of SiO masers is a small fraction of the ring. The above points imply that SiO masers can be used as powerful probes of the processes that drive the mass loss and dynamics in the inner region of the CSEs.

SiO masers can also be  important for understanding the evolutionary path from the spherically symmetric AGB stars to aspherical Planetary Nebulae (PNe). Magnetic fields likely play an important role in shaping the CSEs of evolved stars \citep[e.g.][]{garcia1997}. Polarization observations of circumstellar masers enable us to determine the magnetic field strength and morphology at different distances from the central stars. Observations indicate that SiO masers are significantly linearly polarized, with polarization vectors mainly tangential to the maser ring. For example, multi-epoch polarization VLBA observations of the Mira star TX Cam revealed the linear polarization morphology tangential to the projected shell of SiO maser emission \citep[e.g.][]{kemball2009}. However, Cotton et al. (2008), observed five Mira variables with the VLBA at 43 GHz, but not all sources showed a tangential linear polarization morphology. It remains therefore unclear whether the tangential linear polarization morphology is a generic property of the SiO maser region of evolved stars. The circular polarization of the masers is in the range 3$\%$-5$\%$ \citep[e.g.][]{barvainis1987,kemball2009}.
Polarimetric observations of a sample of evolved stars indicate an average magnetic field of several gauss \citep[][]{herpin2006,kemball1997}. The SiO maser magnetic fields are consistent with those 
measured using H$_2$O and OH masers further out in the envelope assuming a dipole ($B \propto r^{-3}$) or solar-type ($B \propto r^{-2}$) magnetic field  \citep[][]{wouter2005}.

VLBI studies of SiO masers have so far focused on Mira variables and a handful of post-AGB objects and water fountain sources \citep[e.g.][]{cotton2008,desmurs2007,imai2005}, and as far as we can tell no high resolution VLBI observation has been performed to image SiO masers in OH/IR stars. These objects have larger CSEs and much longer periods up to 2000 days compared to those of Mira variables \citep[][]{herman1985}.  They are strong 1612 MHz OH maser emitters \citep[][]{baud1979}. The stars are surrounded by thick dust shells which makes them optically obscured. The SiO maser pump mechanism in OH/IR stars may be different from that operating in Mira variables \citep[][]{doel1995}. Due to the higher mass-loss of these objects the CSE is denser at least an order of magnitude more than those of Mira variables.

Here, we report the SiO maser polarimetric observations of the OH/IR star OH44.8-2.3 with the VLBA. The observations enable us to obtain the spatial distribution of the SiO maser features in OH/IR
stars for the first time. Additionally, our experiment probes the magnetic field strength and morphology in the SiO maser region of OH/IR stars and compares them with those of Mira variables.  We also reduced the 1612 MHz OH maser observations of this star from the VLA archive in order to determine whether there exists any large scale asymmetry in the OH maser shell of this star.

OH~44.8-2.3 is identified as a moderate OH/IR star with a warm and thin CSE and a period of 534 days \citep[][]{groenwegen1994}. The mass loss of this star is estimated to be $\sim$4.6$\times$10$^{-6}$ M$_{\odot}$/yr  \citep[][]{verhoelst2010}. The OH masers of this source show a double peak profile at -88.9 and -53.8 km s$^{-1}$ with a peak flux density of 14.4 Jy for the red shifted part of the spectrum \citep[][]{engels2007}. A distance of 1.13$\pm$0.34 kpc was measured for this star using the phase-lag method based on the  characteristic 1612 MHz double peak profile of the OH masers of this source \citep[][]{langevelde1990}. The H$_2$O masers of this star were detected by \cite{engels1986} at $\sim$5 Jy. However, further single dish monitoring of this source did not result in any detection \citep[][]{shintani2008, kim2010}.
The v=1, $J = 1\to 0$ SiO masers of this star were observed by \cite{nyman1998} with a flux density of 17.6 Jy. \cite{kim2010} observed the v=1, 2 , $J = 1\to 0$ emission from this star with an integrated flux density of $\sim$25 Jy and $\sim$28 Jy, respectively.

The outline of the paper is as follows: the observations are described in Sec.~\ref{observations ch6}. Before giving the results in Sec.~\ref{results ch6}, we introduce the necessary background on maser polarization in Sec.~\ref{sio theory}. We discuss the interpretation of the results in Sec.~\ref{discussion ch6}. This is followed by conclusion in Sec .~\ref{conclusion ch6}.

\section{Observations}\label{observations ch6}
We observed the v=1, $J = 1\to 0$ SiO maser emission toward OH~44.8-2.3 on 6 July 2010 using the NRAO \footnote{The National Radio Astronomy Observatory (NRAO) is a facility of the National Science Foundation operated under cooperative agreement by Associated Universities, Inc.}  Very Long Baseline Array (VLBA) operating in the 43 GHz band. Additionally auxiliary Extended Very Large Array (EVLA) interferometer observations were performed on 2 July 2010 and 11 July 2010 to allow for the absolute calibration of the electric vector polarization angle (EVPA). Furthermore, we retrieved the 1612 MHz OH maser very large array (VLA) observations of OH~44.8-2.3 from the NRAO archive. The VLBA, EVLA and VLA observations are discussed below:

\subsection{VLBA observations and reduction}
The data were recorded in dual circular polarization spectral line mode, which generates all four polarization combinations in the correlator. The DiFX correlator was used with a bandwidth of 4 MHz and 1024 spectral channels, which results in 0.03~km~s$^{-1}$ spectral resolution. The observations were performed in one spectral window centered at a fixed frequency corresponding to the v=1, $J = 1 \to 0$ SiO maser transition at a rest frequency of 43.12208 GHz and a stellar velocity of -72 km s$^{-1}$  with respect to the local standard of rest (LSR), determined from OH maser observations. The total observing time was 6 hr, balanced between the target source, OH~44.8-2.3, and the continuum calibrators, J2253+1608, J1751+0939 and J1800+3848. We achieved a spatial resolution of 0.5$\times$0.2 mas. 

We used the Astronomical Image Processing Software Package (AIPS) to perform the rest of the calibration, editing and imaging of the data. The first calibration steps were performed on the data set with modest spectral resolution (128 channels). The solutions were then applied to the high spectral resolution data set (1024 channels). Parallactic angle corrections were performed on all calibrators. The North Liberty (NL) and Saint Croix (SC) antennas had to be flagged due to the bad weather conditions at the time of the observations.

Since the SiO molecule is non-paramagnetic, the fractional circular polarization will be small ($m_{c} \sim 1\%-3 \%$). In order to preserve the low stokes V signature, we performed amplitude calibration of OH~44.8-2.3 using the AIPS task 'ACFIT'. In this method, each circular polarization autocorrelation spectrum was calibrated independently using the method described by \cite{reid1980}. A template auto correlation spectrum was selected for $\sim$20 min interval from the Los Alamos antenna at a time range with sufficiently high elevation. The template spectrum was fitted to all other total power spectra from all antennas and the relative gains of the antennas as a function of time were determined. We used the system temperature measurements provided with the data to perform the amplitude calibration for the calibrators.

 The complex band pass solutions were obtained using J2253+1608 on each data set separately. Fringe fitting for the residual relay and rate for the parallel hand data was done using the continuum calibrators J2253+1608 and J1751+0939. The fringe rate solutions for OH~44.8-2.3 were determined on the strongest feature. The delay offset between the right and left circular polarizations at the reference antenna was determined for J2253+1608 on the Los Alamos - Fort Davis baseline. The solutions were transferred to all other baselines and sources. The polarization leakage terms were determined using J2253+1608. After the solutions were applied to both data sets image cubes could be created with 50 $\mu$arcsec pixel spacing. The rms noise in the modest spectral resolution and the high spectral resolution individual channel images corresponds to 8 mJy and 26 mJy, respectively.

\subsection{EVLA observations and reductions}
The absolute phase difference between the right and left circular polarizations at the reference antenna is not known, which implies that the absolute EVPA of linearly polarized emission cannot be measured accurately in our VLBA observations. Therefore, auxiliary EVLA observations were performed to measure the absolute polarization angle of the polarization calibrators with respect to a primary polarization calibrator with known polarization angle. 

The continuum sources J2253+1608 and J1751+0939 were observed as transfer calibrators on 2 July 2010 and 11 July 2010, respectively. The observations were performed in continuum mode and full polarization, using two 128 MHz spectral windows. The data reduction was performed for each spectral window separately. We achieved a beam size of 2.6$\times$1.7 arcsec during the observations.

We reduced the data according the standard EVLA data calibration and imaging recipe in the appendix of AIPS Cook Book. Since the delays are not set accurately in EVLA observations, we solved for the phase slope using the AIPS task FRING. A $\sim$1 min solution interval was chosen on J2253+1608 and J1751+0939 to solve for the phase slope as a function of frequency. The bandpass solutions were determined using J2253+1608 and J1751+0939. The flux density of the primary flux calibrators 3C286 and 3C48 was determined using the model provided within the AIPS software. The AIPS task 'CALIB' was used to perform phase calibration for 3C286 and 3C48 using a model of the sources. Subsequently, the phase solution was determined for the secondary calibrators J2253+1608 and J1751+0939. The amplitude of the secondary calibrators was adjusted with respect to the flux density of the primary flux calibrator. We used 3C286 and 3C48 for polarization calibration. The delay difference between the right and left circular polarizations were determined using the AIPS task 'RLDLY'. The feed parameters for each polarization were determined both in continuum and spectral mode since the EVLA feeds have significant variations in frequency. Images were produced in each spectral window in Stokes Q, U, I adopting a pixel spacing of 0.2 arcsec. The rms noise for J2253+16 and  J1751+9839 were $\sim$8 mJy and $\sim$1 mJy, respectively.

\begin{figure*}
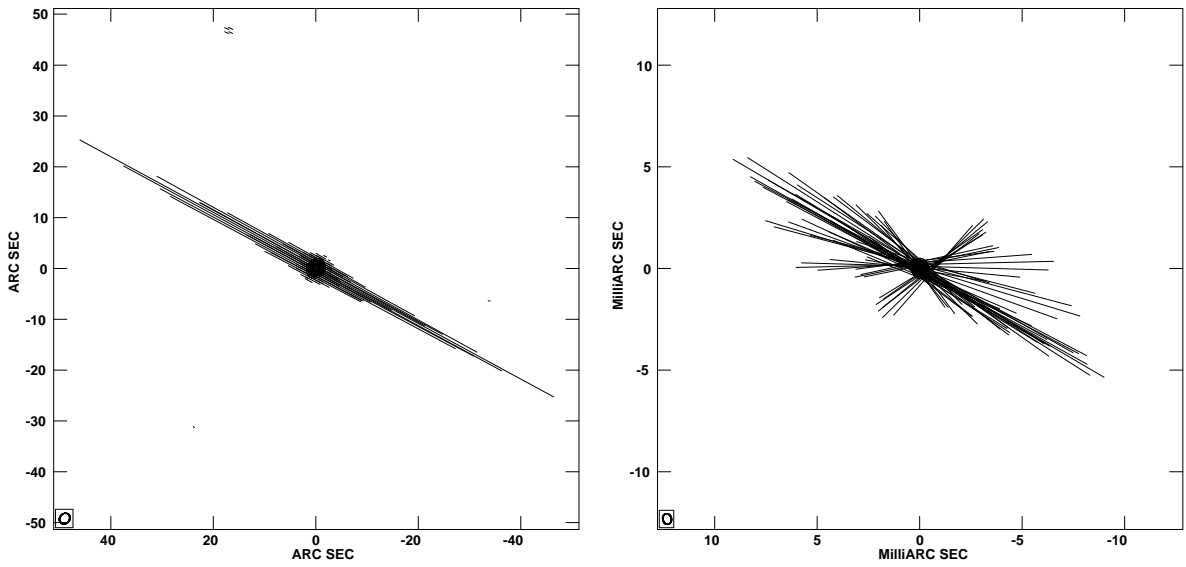

\center
\includegraphics[scale=0.4,angle=-90]{./J22_evla.ps}
\includegraphics[scale=0.4,angle=-90]{./j22_vlba_pcntr.ps}
\caption{ Linearly polarized map of the polarization calibrator J2253+16 obtained with the EVLA (left panel) and the VLBA (right panel). The vectors are rotated by 15$^{\circ}$ in the VLBA image.}
\label{fig:j2253}
\end{figure*}

The absolute EVPA of J2251608 and J1751+0939 were measured with respect to 3C48 and 3C286, respectively. The final absolute EVPA was calculated as the mean of the EVPA in each spectral window. This corresponds to the measured EVPA of 62$^{\circ}$ and -44$^{\circ}$ for J2253+1608 and J1751+0939, respectively. Fig. \ref{fig:j2253} shows the EVPA of the linearly polarized emission for J2253+1608 for both EVLA and VLBA observations. The EVPA in the VLBA image is rotated by 15$^{\circ}$, which corresponds to the difference in the EVPA between the EVLA and VLBA images. We note that the linearly polarized emission of the other polarization calibrator, J1751+0939 shows two polarized components separated by $\sim$90$^{\circ}$ in the VLBA image. Therefore we were not able to use this source as a robust EVPA calibrator.

\subsection{VLA observations of the 1612 MHz OH masers of OH~44.8-2.3}
We found the previous observations of the 1612 MHz OH masers of OH~44.8-2.3 observed on 25 September 1983 using the Very Large Array (VLA) from the NRAO archive. The observations were performed under the project name 'AH127' in the A configuration with the largest spacing of 36.5 km, which gives a resolution of 1$''$. The band width of 1.5 MHz was used for the observations which gives a velocity resolution of 1.4 km s$^{-1}$ at 1612 MHz. 3C286 was observed as a flux density calibrator and 1741-038 was used as the secondary phase calibrator. The data reduction was performed following the recipe provided in the AIPS cookbook.

\begin{figure*}
\center
\includegraphics[scale=0.33,angle=0]{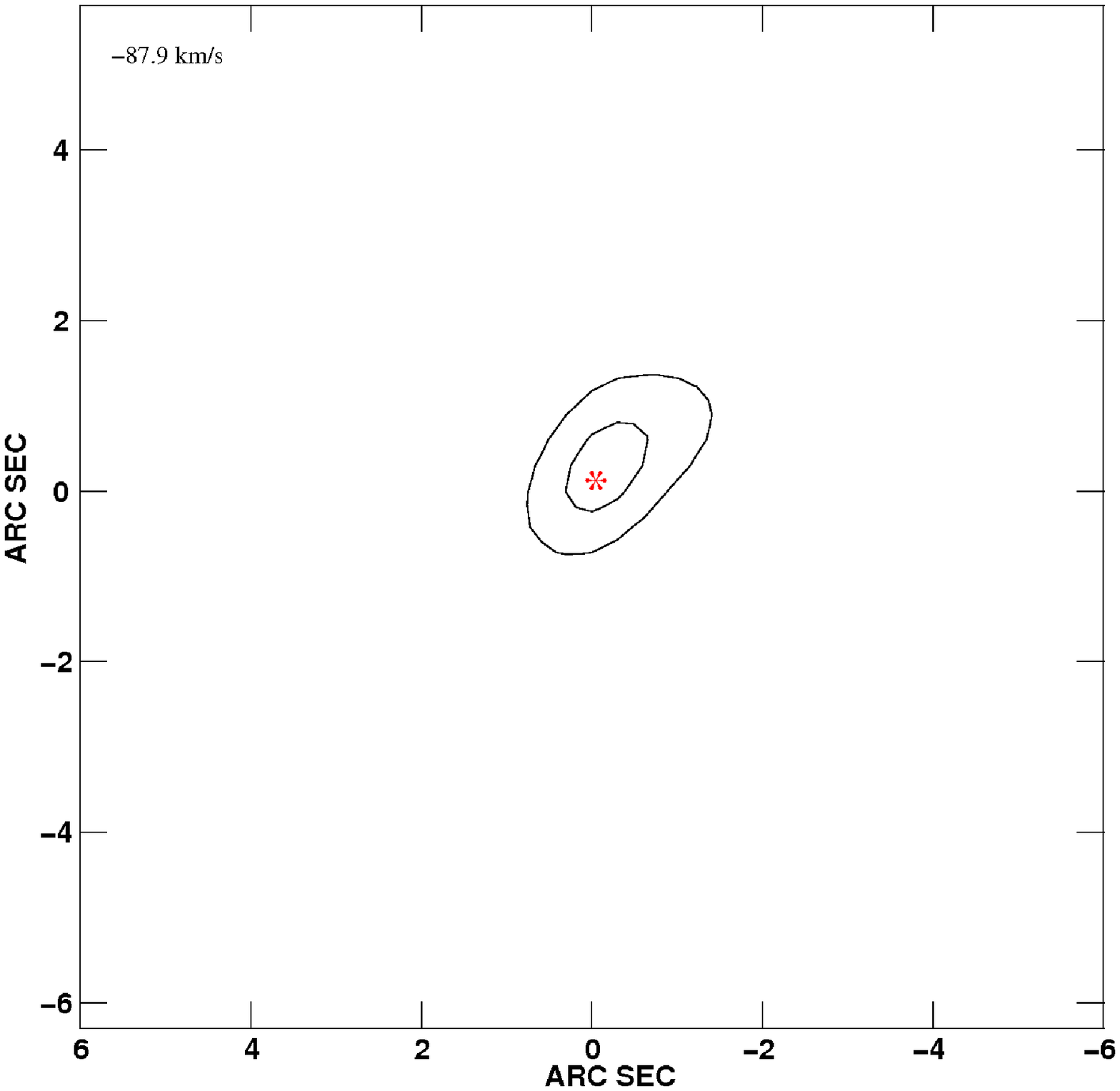}
\includegraphics[scale=0.33,angle=0]{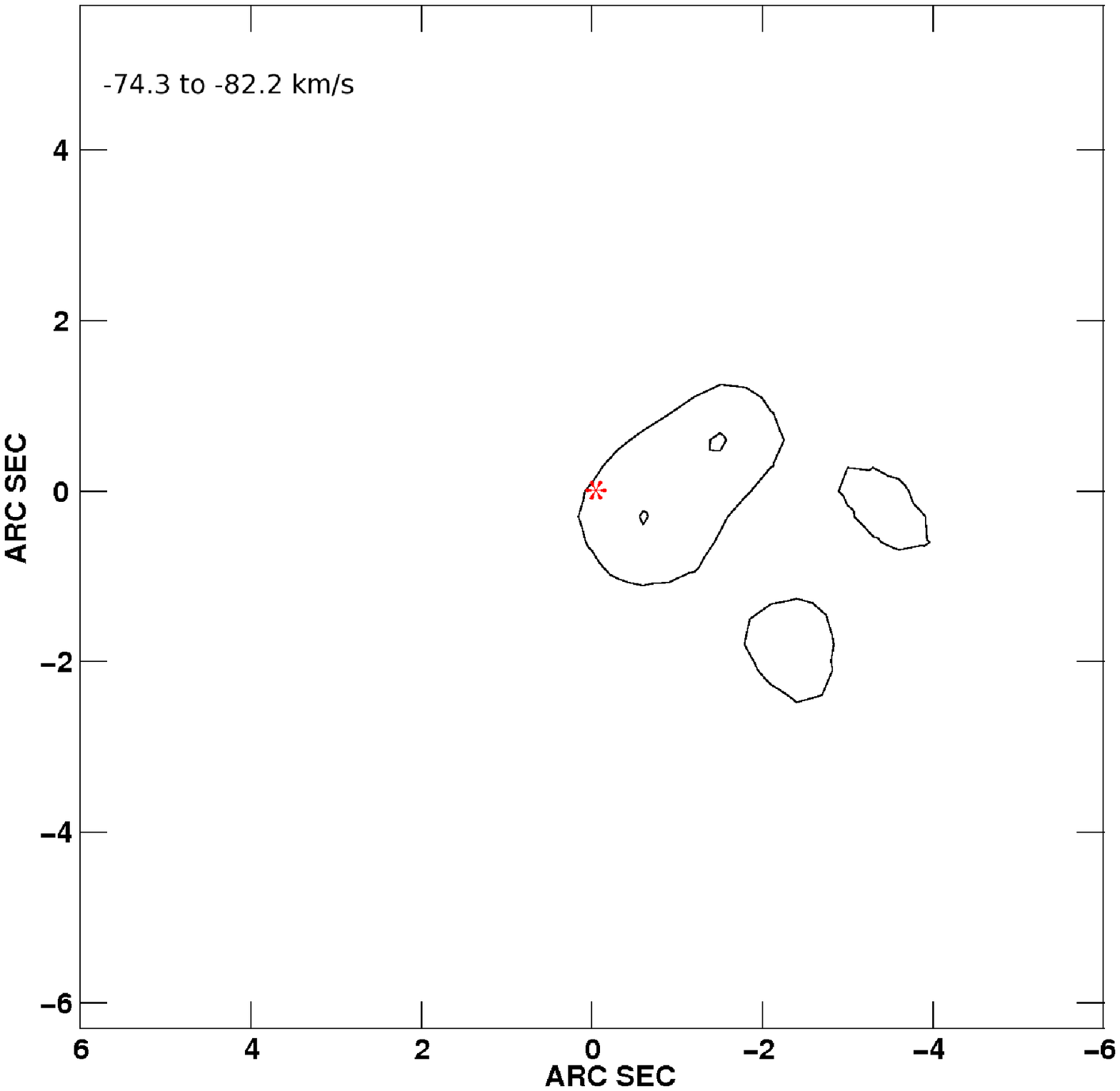}
\includegraphics[scale=0.33,angle=0]{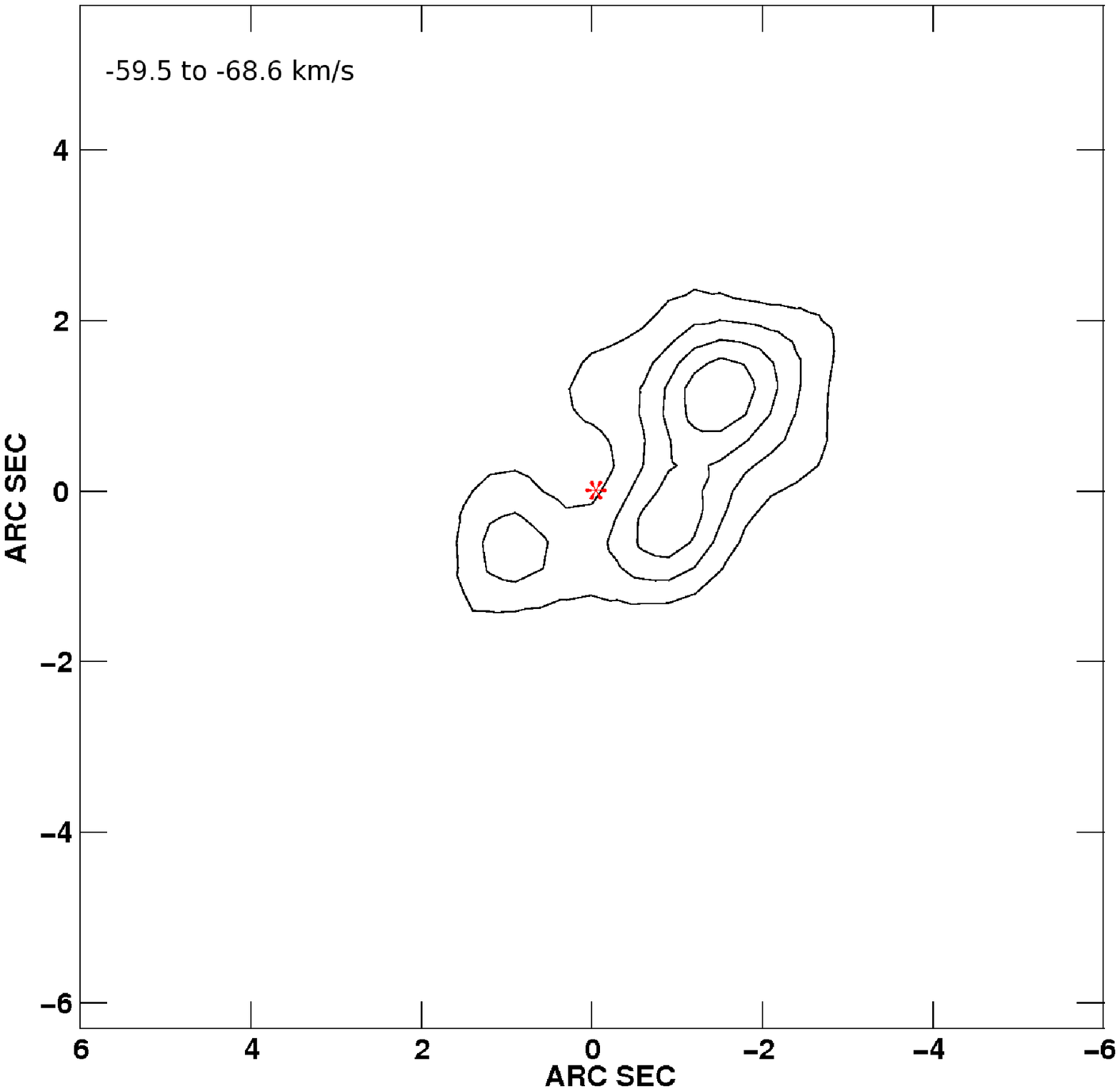}
\includegraphics[scale=0.33,angle=0]{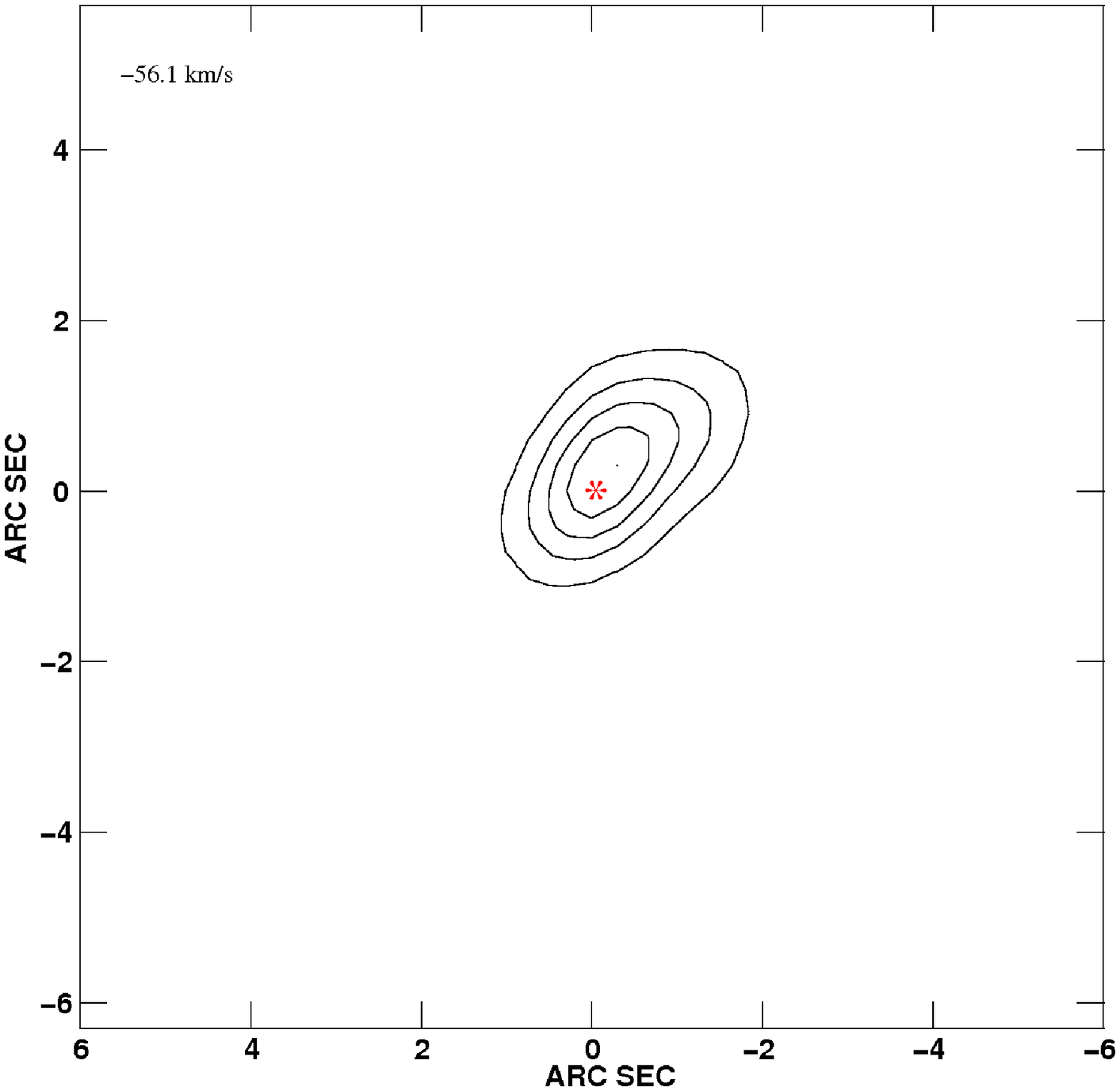}
\caption{The 1612 MHz OH maser map of OH~44.8-2.3 obtained with the VLA. The top-left and bottom-right panels show the blue- and red-shifted peaks. The red-shifted peak exhibits the highest flux density of 8.1 Jy. The contour levels are at 1.62, 3.24, 4.86, 6.48 and 8.1 Jy. The top-right and bottom-left panels exhibit emission summed over several channels close to the stellar velocity. The contour levels are at 0.005, 0.011, 0.016, 0.022 and 0.027 Jy. The star symbol indicates the position of the peak in the red-shifted emission which likely indicates the position of the central star.}
\label{fig:OH1612_map}
\end{figure*}

Fig. \ref{fig:OH1612_map} displays the OH maser map of OH~44.8-2.3. The emission covers a velocity range of -87 to -56 km s$^{-1}$. The figure shows the blue- and red-shifted peaks as well as emission summed over several channels close to the stellar velocity.

\section{SiO maser polarization theory}\label{sio theory}

The SiO maser emission at 43.12208 GHz involves the rotational transition $J = 1\to 0$ in the excited vibrational state v=1. Since the SiO molecule is non-paramagnetic, its response to the magnetic field is weak. This implies that the Zeeman splitting is smaller than the line width. 

\subsection{Linear polarization}
SiO masers present high fractional linear polarization \citep[e.g.~][]{kemball1997}. It was shown that anisotropic background radiation from the central star generates anisotropic pumping, potentially producing a high degree of linear polarization in the SiO emission region \cite[e.g.][]{watson2002}. In the case where the SiO masers are radiatively pumped, the magnetic sub states may be anisotropically populated, which can generate high linear polarization fraction for saturated masers \cite{western1983}. However, the results from \cite{nedoluha1990} show that the polarization vectors still trace the direction of the magnetic field despite the fact that the linear polarization may originate largely from anisotropic pumping.

\subsection{Potential non-Zeeman effects for circular polarization}

The  non-Zeeman mechanism prohibits one to interpret circular polarization as a measure of the magnetic field strength. This stems from the competition between the stimulated emission rate (R), the Zeeman coefficient rate (g$\Omega$) and the radiative decay rate ($\Gamma$). The circular polarization can be created by the change of the axis of symmetry for the molecular quantum states when the condition g$\Omega \simeq R > \Gamma$ is satisfied  \citep[][]{nedoluha1994}. For some distance along the maser path g$\Omega \ge R$ is satisfied and the magnetic field is the quantization axis. As the radiation propagates farther into the maser path and the rate of stimulated emission becomes larger, the inequality  R $\ge$ g$\Omega$ is satisfied and the molecule interacts more strongly with the radiation. This implies that the axis of the symmetry of the molecule changes from parallel to the magnetic field to parallel to the direction of propagation. The resulting circular polarization would have the anti-symmetric profile which resembles to that produced by the ordinary Zeeman effect. The intensity dependent circular polarization can be  higher than that created by the Zeeman effect by a factor as large as 1000. This scenario is an inherent part of the radiative transfer process and does not require anisotropic pumping or high fractional linear polarization.

\cite{wiebe1998} introduce yet another non-Zeeman effect in which the propagation of a strong linear polarization can create circular polarization if the condition  $g\Omega \gg R > \Gamma$ is satisfied. The circular polarization can be generated if the magnetic field orientation changes along the maser propagation direction. The circular polarization produced from this scenario is on average $\sim \frac{m_{l}^{2}}{4}$ where $m_{l}$ indicates the linear polarization fraction. For a typical linear polarization fraction around 30$\%$, the circular polarization results from the formula above corresponds to $\sim$2-3$\%$  which is in agreement with the observed circular polarization \citep[e.g.][]{barvainis1987}, however in individual maser features the circular polarization fraction can go up to $\sim$20$\%$. This mechanism would then generate circular polarization with only 10 - 20 mG fields. Moreover, if there is no velocity gradient along the maser path, the circular polarization profile would have the shape of the S-curve. However, the intrinsic profiles can be distorted as a result of saturation, together with velocity and magnetic gradients along the amplification path. A correlation between the linear and circular polarization fraction is expected for the individual features.

\section{Results}\label{results ch6}
\subsection{Total intensity}
\begin{figure*}
\center
\includegraphics[scale=0.8,angle=0]{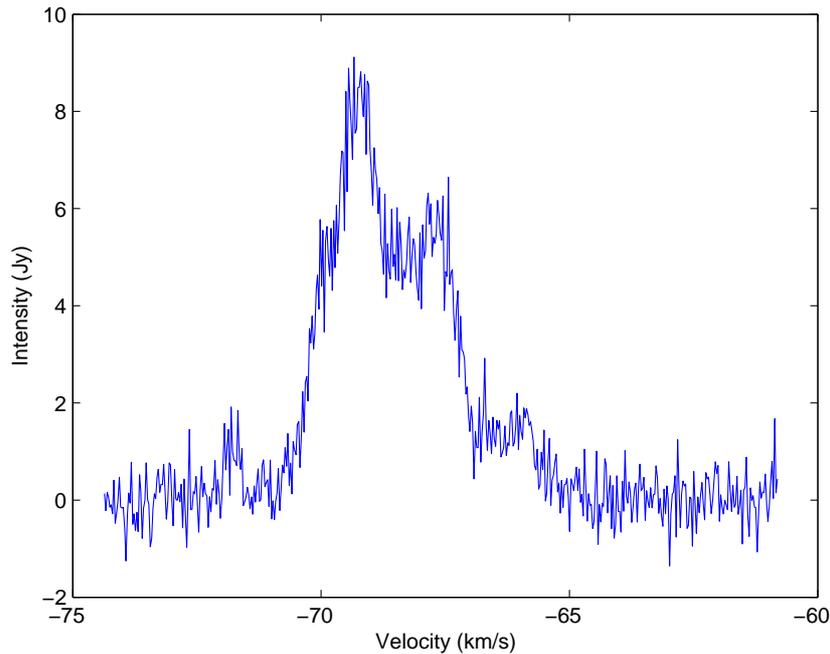}
\caption{The total intensity spectrum of OH~44.8-2.3 for the v=1, $J = 1\to 0$ SiO maser emission obtained with the VLBA.}
\label{fig:OH44_spec}
\end{figure*}

\begin{figure*}
\center
\includegraphics[scale=0.9,angle=0]{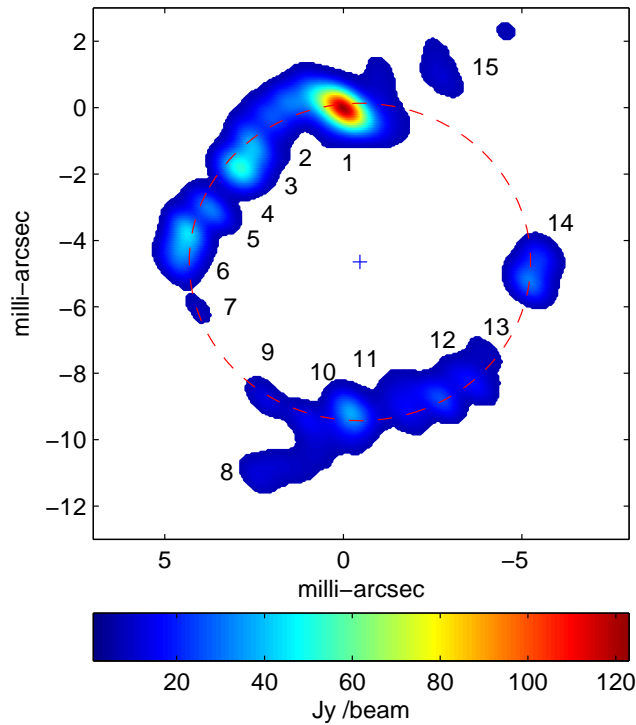}
\caption{The VLBA map of v=1, $J = 1 \to 0$ SiO maser emission towards the OH/IR star OH~44.8-2.3. The features are color coded according to the flux density (Jy / beam) integrated over all velocity channels.}
\label{fig:OH44_ring}
\end{figure*}

Fig. \ref{fig:OH44_spec} shows the total intensity spectrum of the SiO maser emission towards OH~44.8-2.3. The emission covers a velocity range of $\sim$7 km s$^{-1}$. The SiO emission 
actually implies a stellar velocity ~3-4 km s$^{-1}$ offset from the stellar velocity of -72 km s$^{-1}$ based on OH maser observations of this star. Fig. \ref{fig:OH44_ring} displays the SiO maser emission map of OH~44.8-2.3 summed over all velocity channels covering emission. The maser features form a partial ring of 4.75 mas corresponding to 5.4 AU assuming a distance of 1.13 kpc \citep[][]{langevelde1990}. The masers appear to be absent from the North-East and South-West part of the ring. The ring pattern observed in the SiO maser region of this star implies that the masers are tangentially amplified. This is typically interpreted to indicate that large velocity gradients exist in the SiO maser region which prohibits radial amplification. The peak flux density for the maser features are displayed in Table \ref{tab:sio results}. Feature 1 exhibits the largest flux density of 2.8 Jy.

\begin{table*}[htdp]
\begin{center}
\begin{tabular}{|c|c|c|c|c|c|c|}
\hline
feature	&	RA	&	Dec	&	V	&	I	&	m$_l$	&		B 	\\	
       &	19 21 &  09 27   &   km s$^{-1}$   & Jy& $\%$   & G \\
\hline
1	&	36.63700337	&	56.5199500	&	-69.4	&	2.9	&	8	& 1.5 $\pm$ 0.3	\\
2	&	36.63710476	&	56.5203000	&	-69.8	&	0.7	&	26	&	$<$	3	\\
3	&	36.63722303	&	56.5180500	&	-67.8	&	0.5	&	23	&	$<$	9	\\
4	&	36.63725683	&	56.5169000	&	-65.9	&	0.9	&	36	&	$<$	3	\\
5	&	36.63728386	&	56.5161500	&	-67.7	&	1.2	&	34	&	$<$	2	\\
6	&	36.63731427	&	56.5152000	&	-68.3	&	0.7	&	48	&	$<$	2.5	\\
7	&	36.63727710	&	56.5141000	&	-66.6	&	0.3	&	50	&	$<$	3	\\
8	&	36.63717234	&	56.5091500	&	-70	&	0.6	&	100	&	$<$	2.7	\\
9	&	36.63715545	&	56.5113500	&	-68.7	&	0.4	&	50	&	$<$	12	\\
10	&	36.63706420	&	56.5106000	&	-71.2	&	0.6	&	42	&	$<$	3	\\
11	&	36.63698310	&	56.5111500	&	-67.2	&	0.6	&	18	&	$<$	36	\\
12	&	36.63680399	&	56.5109500	&	-67.7	&	0.5	&	30	&	$<$	6	\\
13	&	36.63673641	&	56.5116000	&	-67.9	&	0.3	&	69	&	$<$	7	\\
14	&	36.63661475	&	56.5156000	&	-69.0	&	0.7	&	46	&	$<$	2.5	\\
15    &	36.63682427	&	56.5214500	&	-69.2	&	0.24	&	-	&	$<$ 12\\

\hline
\end{tabular}
\end{center}
\caption{ Results of the Magnetic fields determination of the SiO maser features of OH~44.8-2.3 obtained from the VLBA observations. }
\label{tab:sio results}
\end{table*}%

\subsection{Linear polarization}

\begin{figure*}
\center
\includegraphics[scale=0.5,angle=-90]{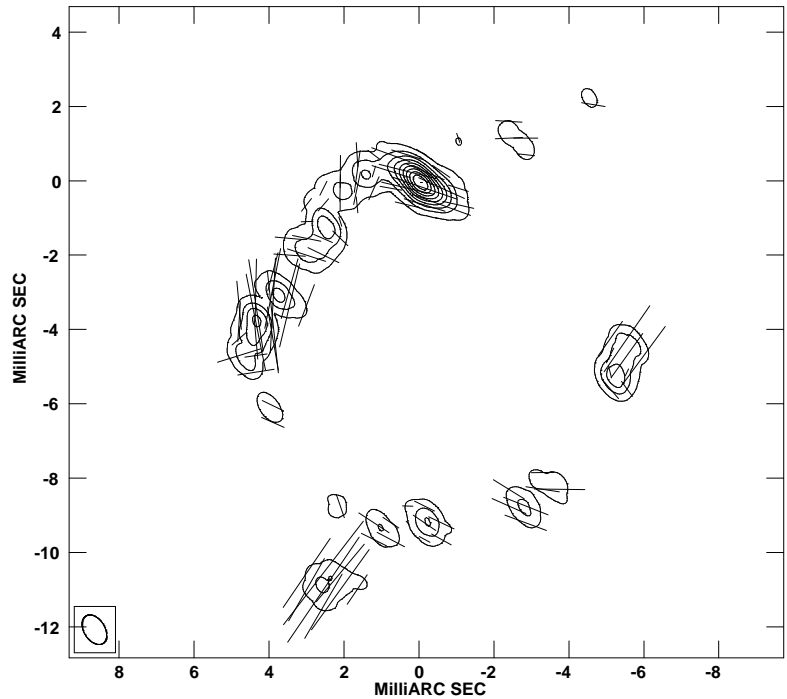}
\caption{Contour plot of the Stokes I image at levels [1, 2, 5, 10, 20, 40, 80, 100]$\%$ of the peak. Vectors are overlaid proportional to the linearly polarized intensity (on a scale 1 mas = 1.25 Jy beam$^{-1}$) and drawn at a position angle of the EVPA. All Stokes parameters (I,Q,U) are summed over
velocity.}
\label{fig:OH44_linpol}
\end{figure*}

The polarization morphology is shown in Fig. \ref{fig:OH44_linpol}. In this plot, the polarized emission is plotted as vectors with a length proportional to the polarization intensity. The position angle of the vectors corresponds to the EVPA of the emission. All stokes parameters are summed over frequency before making this plot. The maps of stokes I, Q, and U are available via CDS. The background contours represent the total intensity image. The fractional linear polarization for the individual maser features is shown in Table \ref{tab:sio results}. The average linear polarization fraction for the SiO maser features of this star is $\sim$30$\%$. In particular for feature 8 it reaches 100$\%$. This likely implies that the (radiative or collisional) pumping of SiO masers is not isotropic.

\subsection{Circular  polarization}\label{sec:circular fraction}

\begin{figure*}
\center
\includegraphics[scale=0.8,angle=0]{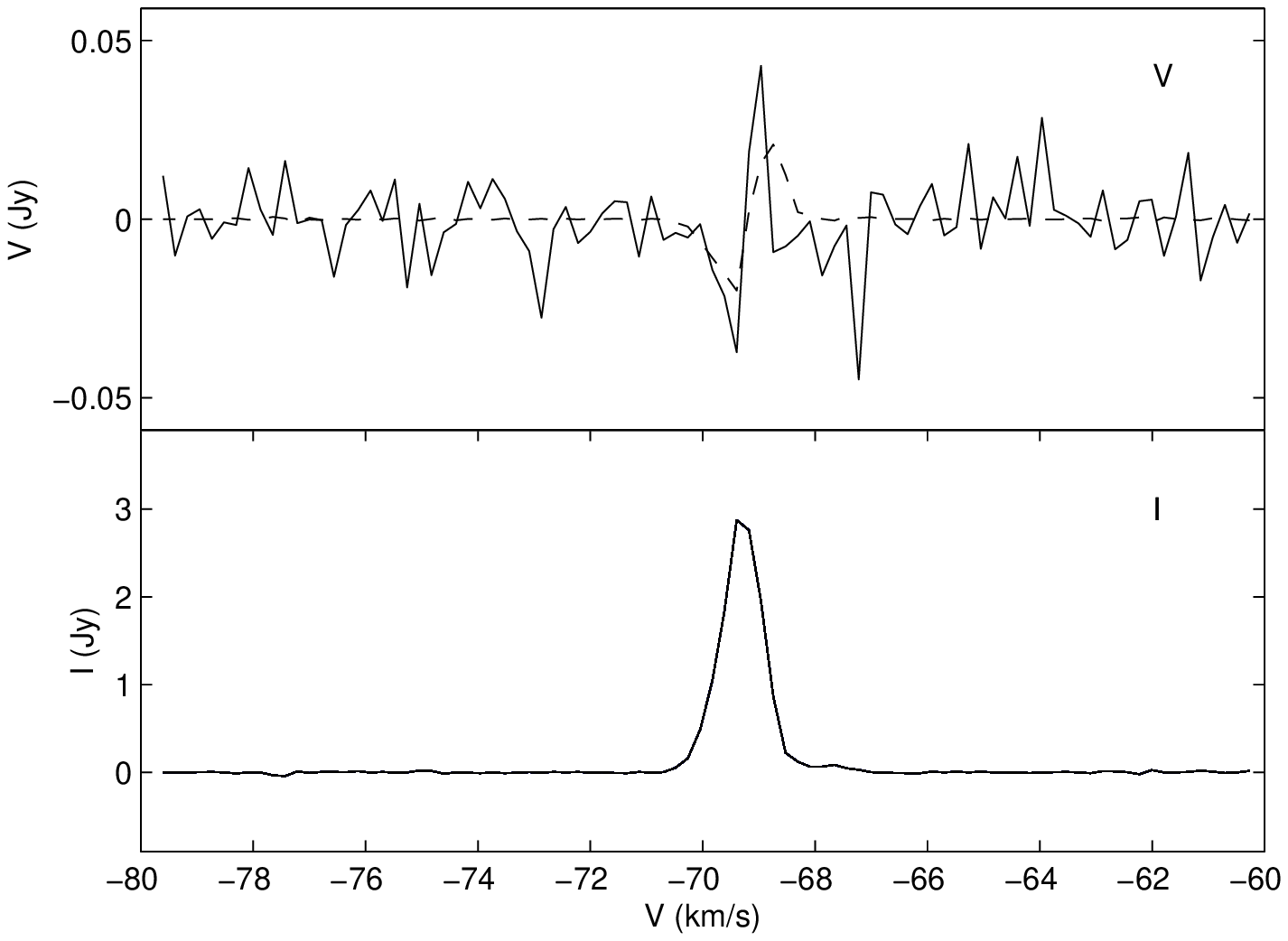}
\caption{Total power (I) and circular polarization (V ) spectra of the brightest SiO maser feature of OH~44.8-2.3. The dashed line is the fit to the observed V-spectrum . The V spectrum is shown after removing the scaled down replica of stokes I.}
\label{fig:circular sio}
\end{figure*}

The instrumental gain differences between the RCP and LCP spectra, can cause a residual scaled down version of the Stokes I spectrum to appear in the circular polarization profile. The Zeeman splitting for the stokes V spectra of the SiO maser features of OH~44.8-2.3 can be determined by measuring the residual curve having the shape of the derivative of the total intensity spectrum in the least square sense \citep[][]{troland1982}:

\begin{equation}\label{eq:v}
V(\nu)=a\times  \frac{dI}{d\nu} + b \times  I(\nu)
\end{equation}
Where a is the measure of the circular polarization due to the Zeeman splitting and b denotes the instrumental differential amplitude. This method assumes that intrinsically equal flux is present in the RCP and LCP spectra, which implies that the circular polarization spectra have the antisymmetric S-curve pattern. However, theoretical models of circular polarization produce asymmetric profiles depending on the velocity gradient along the amplification path \citep[][]{wiebe1998}. We find a tentative detection of the circular polarization for the brightest maser feature in Fig. \ref{fig:OH44_ring} (feature 1 from Table \ref{tab:sio results}). Fig. \ref{fig:circular sio} displays the circular polarization together with the total intensity spectra for this feature. The figure also shows the fit to the circular polarization profile after removing the scaled down replica of stokes I. The circular polarization fraction is measured as:
\begin{equation}
m_{c} = \frac{V_{max}} {I_{max}} \times 100
\end{equation}
Where V$_{max}$ corresponds to the maximum of the fit to the observed stokes V spectrum (Eq. \ref{eq:v}: $a \times \frac{dI}{d\nu}$). For the modest spectral resolution data (128 channels), we measure  circular polarization of  $m_{c}\sim0.7 \pm 0.2\%$. However, due to the increased noise in individual channels in the high  spectral resolution data, we can not confirm the detection. The magnetic field derived from circular polarization corresponds to the following equation \citep[][derived from Elitzur (1996)]{kemball1997}:
\begin{equation}
B=3.2 \times m_{c} \times \Delta\nu_{D} \times \cos\theta
\end{equation} 
Where m$_{c}$, $ \Delta\nu_{D}$ and $\theta$ indicate the fractional circular polarization, the maser line width and the angle between the magnetic field and line of sight, respectively. The full width half maximum line with for the stokes I spectrum of feature 1 corresponds to $\sim$ 0.8 km s$^{-1}$. Using the preceding relation a magnetic field of 1.8$\pm$0.5 G is derived for feature 1.

\subsubsection{Cross-correlation method and magnetic field strength}\label{sec:crosscorr}

Alternatively, we use the cross-correlation method introduced by \cite{modjaz2005} to measure the magnetic field due to the Zeeman splitting. In this method the right circular polarization (RCP) and the left circular polarization (LCP) spectra are cross-correlated to determine the velocity splitting.  The magnetic field is determined by applying the Zeeman splitting coefficient for SiO masers. This method can even work for complex spectra, assuming that the velocity shift is the same over the spectrum; which means the magnetic field strength and direction is constant in the maser region. The sensitivity of this method is comparable to the S-curve method, where the stokes V spectrum is directly used for measuring the magnetic field \citep[e.g.][]{wouter2001,fiebig1989}.

The Zeeman rate for SiO is given by \citet{nedoluha1990} as:

\begin{equation}\label{eq:zeeman}
\frac{g\Omega}{2\pi} \sim 200 \left(\frac{B}{\rm{G}}\right)\ s^{-1}\label{eqn-go}
\end{equation}
The equation above implies the Zeeman splitting coefficient for the $J = 1\to 0$, v=1 transition of the SiO maser  as 8.7$\times$10$^{-3}$~ km~ s$^{-1}$~ G$^{-1}$.  We present the result of the cross-correlation method in Table \ref{tab:sio results}. We measured a magnetic field of 1.5$\pm$0.3 G for feature 1.  However, for the other features of the SiO emission of this star we only place upper limits at 3$\sigma$ level.

\subsection{OH maser observations of OH~44.8-2.3}

The VLA results on OH 1612 MHz emission from OH~44.8-2.3 were not presented directly in the scientific literature before, although the original distance determination in \cite{herman1986} and the one we use from \cite{langevelde1990} depend directly on it. We have access to the results of the original analysis in which a shell radius of 1.302$"\pm0.023"$ was obtained (van Langevelde, private communication).  The new processing we performed gave consistent images with these previous results, but we note that the OH at intermediate velocities is quite elongated (Fig. \ref{fig:OH1612_map}). The inferred size of the region is of course much larger than the SiO, spanning $\sim$1471 AU compared to 5.4 AU. One should note that with typical outflow velocities of $\sim$16 km s$^{-1}$ and 5 km s$^{-1}$ for OH and SiO masers, the dynamical timescales are also much different: $\sim$440 years for the OH and $\sim$5 years for the SiO.

\section{Discussion}\label{discussion ch6}
\subsection{Linear polarization}

In order to the compare the two conflicting polarization theories (a magnetic versus anisotropic pumping origin of the linear polarization implying a Zeeman or non-Zeeman origin of the circular polarization), we need to determine in which theoretical regime the masers were observed. This includes the determination of the stimulated emission rate (R), the Zeeman rate ($g\Omega$), and the collisional and radiative decay rate $\Gamma$. The  radiative decay rate for the v=1, $J=~1~\to~0$ transition of the SiO molecule is estimated to be 5 s$^{-1}$ \citep[][]{kwan1974,elitzur1992}. The stimulated emission rate (R) is estimated by \citep[][]{kemball2009}.

\begin{equation}\label{eq:R}
R=23\ \left(\frac{T_B}{2\times10^{10}\ \rm{K}}\right)\ \left(\frac{d\Omega}{10^{-2}\ \rm{sr}}\right) \rm{s}^{-1}\label{eqn-r}
\end{equation}

Where $T_B$ is the maser brightness temperature and $d\Omega$ is the estimated maser beaming angle. We obtained an estimate for the brightness temperature for feature 1 (Fig. \ref{fig:OH44_ring}) as 5$\times$10$^{10}$ K. We adapt an estimated beaming angle of 10$^{-2}$ sr \citep[][]{kemball2009}. Replacing these values in eq. \ref{eqn-r} results in R=57.5 s$^{-1}$. Substituting the value of $\sim$1.5 G obtained for feature 1 (Table \ref{tab:sio results}) in eq. \ref{eqn-go}, implies $g\Omega \sim$1900 s$^{-1}$. Therefore, we are in a regime where $g\Omega \gg R  >  \Gamma$. 
In this regime the linear polarization vectors appear either parallel or perpendicular to the projected magnetic field, depending on the angle between the magnetic field direction and the line of sight  \citep[][]{goldreich1973}. Taking the EVPA of -50$^{\circ}$ for feature 8 which has the highest linear polarization fraction, this implies that the magnetic field direction is either parallel ($-50^{\circ}$) or perpendicular (40$^{\circ}$) to the linear polarization vectors. In either cases, the EVPA vectors likely indicate a large scale magnetic field in the SiO maser region of this star.

The linear polarization morphology of this star is not consistent with the dominant tangential polarization morphologies that were reported for other evolved stars \citep[e.g. TX Cam; ][]{kemball2009}. Based on previous observations of the SiO maser region of TX Cam \citep[][]{kemball1997,desmurs2000}, they conclude that the tangential polarization morphology could be an inter cycle property of the SiO maser emission toward this star. 
Instead of tangential polarization morphology, the EVPA vectors could indicate a bipolar magnetic field morphology for the SiO maser region of OH~44.8-2.3 (Fig. \ref{fig:OH44_linpol}).  However, we can not rule out the other, more complex, morphologies. Polarimetric observations of the OH and H$_2$O masers of this star are required to clarify the magnetic field morphology of this star.
A dipole field morphology is reported for the supergiant VX Sgr using the polarimetric observations of high frequency SiO masers \citep[][]{wouter2011}. This morphology was consistent with the the dipole magnetic field inferred from H$_2$O and OH maser regions at much larger distances from the central star \citep[][]{szymczak1997,szymczak2001,wouter2005}.

\subsubsection{The effects of anisotropic pumping}

The SiO maser features of this star exhibit high fractional linear polarization up to 100$\%$ (Table \ref{tab:sio results}), which likely indicates that the pumping of the masers is highly anisotropic. Such high linear polarization fractions were reported previously \citep[][]{kemball1997,cotton2006}. The masers occur close to the central star and the infrared radiation incident on the masering region is highly anisotropic. This implies that the magnetic sub states are pumped unequally. Additionally, the medium in which they are located likely has large velocity gradients, which leads to preferred direction for the propagation of masers \citep[][]{watson2002}. \cite{western1983} show that even for collisionally pumped SiO masers, the magnetic sub states are anisotropically pumped and high linear polarization would be expected for saturated and unsaturated masers. In the absence of magnetic fields, collisional pumping causes linear polarization along the radial direction. In contrast, radiative pumping would produce linear polarization vectors tangential to the SiO maser ring. However, our observations show that magnetic fields are present in the SiO maser ring of OH 44.8-2.3, which implies that both magnetic field and anisotropic pumping influence the linear polarization morphology. Therefore, we can not distinguish between the radiative or collisional pumping from the observations.

\subsubsection{Maser saturation}
The degree of saturation is the ratio of the rate R for stimulated emission to the loss rate $\Gamma$.
From the observations we measured the stimulated emission rate for feature one as R$\sim$57.5 s$^{-1}$ which implies $\frac{R}{\Gamma} \sim 10$. This indicates that the medium is marginally saturated. In contrast, a saturation of $\frac{R}{\Gamma} \sim 30$ is required to achieve 70$\%$ linear polarization fraction in the non-Zeeman scenario \citep[][]{watson2001}. In the standard Zeeman interpretation scenario proposed by \cite{elitzur1996} the polarization solution does not depend on the saturation level.  \cite{nedoluha1994} show that the low level circular polarization observed for SiO masers can be due to the saturation of masers.

\subsubsection{Jet-like features}

Even though the majority of the SiO maser features  of OH~44.8-2.3 are coherent and fit the ring model, features 8 and 15 do not conform with the ring pattern. A striking characteristic of feature 8 is the 100$\%$ linear polarization fraction. Additionally, the linear polarization vectors of this feature are in the perpendicular direction of the overall linear polarization  morphology of this star (Fig. \ref{fig:OH44_linpol}). \cite{cotton2006} report structures aligned with the direction of the photosphere with magnetic field morphology along the features. They refer to them as jet-like features. They explain that these features likely form due to the masering region being dragged along the magnetic field. For the case of OH~44.8-2.3, the 100$\%$ linear polarization fraction could imply that this feature is a dynamic part of the envelope. Observations of this feature close in time to study the proper motion is necessary to understand the nature of this feature.

\subsection{Circular polarization}

We report a tentative detection of $\sim$0.7$\%$ for the circular polarization fraction for feature 1 (Table \ref{tab:sio results}). The measured magnetic field corresponds to 1.5$\pm$0.3 G for this feature. We note that the magnetic field measured from the circular polarization fraction (Sec. \ref{sec:circular fraction}:1.8$\pm$0.5 G) and the one measured from the cross-correlation method (Sec. \ref{sec:crosscorr}: 1.5$\pm$0.3 G) are consistent within errors. Therefore the correspondence of the field measured from the cross-correlation method with the field determination through the circular polarization fraction confirms that the cross-correlation method works properly.The non-Zeeman effect due to intensity dependent circular polarization introduced by \cite{nedoluha1994} is ruled out since  $g\Omega \gg R$ in our observations. However, since we are in a regime where  $g\Omega \gg R >  \Gamma$ the non-Zeeman effect introduced by \cite{wiebe1998} is applicable.  For the 7$\%$ linear polarization fraction measured for feature 1 (Table \ref{tab:sio results}), the generated circular polarization due to this effect is 0.12$\%$. This implies that the measured circular polarization for feature one is about 6 times higher than the estimated value from the non-Zeeman effect. \cite{wiebe1998} show that if the circular polarization is higher than the average of $\frac{m_{]}^{2}}{4}$, the circular polarization stems from other effects, probably due to the Zeeman effect. Therefore, it is likely that the circular polarization of this star originates from the Zeeman splitting. However, \cite{wiebe1998} explain that the average circular polarization in individual features can go  up to 20$\%$.
Additionally, the determination of Zeeman or non-Zeeman effect from the observed circular polarization profile remains inconclusive, since both models predict the similar anti-symmetric S-curve profile.

\subsection{CSE morphology and magnetic field}
The SiO maser features of OH~44.8-2.3 exhibit two opposite arcs (Fig. \ref{fig:OH44_ring}). Additionally, the OH masers of this star show that the masers have an elongated shell morphology in the direction where there is a gap in the SiO maser emission  (Fig. \ref{fig:OH1612_map}) . We note that the OH masers occur on much larger scale ($\sim$1471 AU) than the SiO masers (5.4 AU) around the star. It is therefore not obvious that both deviations from symmetry are related, but if they are, this is an indication that there is a mechanism at work that can support the asymmetry on many scales. Similar gaps or opposite arcs were seen recently in monitoring observations of the SiO masers of R Cas with the VLBA \citep{assaf2011}. Of course, only with one epoch of observations we can not determine whether the SiO maser arcs in OH 44.8-2.3 are consistent in time since the region where SiO masers occur is expected to be highly dynamic. Therefore, multi-epoch observations of the SiO masers of this star is essential to probe the motion of the SiO maser features.

As mentioned earlier, the magnetic field morphology for the SiO maser region of this star is either parallel or perpendicular to the linear polarization vectors. Interestingly, the direction of the magnetic field is parallel or perpendicular to the location of the gaps in the SiO maser ring and the OH maser extent. There thus appears to be a global preferred direction of the outflow imposed by the magnetic field in the CSE of this star. 
Such asymmetric outflows have already been reported for the CSE of evolved stars. For example, collimated H$_2$O maser jets have been observed in a class of post-AGB objects, the so-called water fountain sources \citep[][]{imai2002,boboltz2005}. Additionally, interferometric observations of the OH maser region of this class of objects revealed aspherical morphologies with either equatorial or bi-conical distributions \citep[][]{amiri2011}. High resolution observations of the SiO masers of the post-AGB object W43A indicated a biconical outflow \citep[][]{imai2005}. The bi-polar jets observed in water fountain sources are related to the onset of asymmetric PNe \citep[][]{sahai1998}.

 The high fractional linear polarization observed for the SiO masers of OH~44.8-2.3 together with the tentative detection of circular polarization is potentially an important indication that magnetic fields have a significant role in shaping the circumstellar environment of this star.
Significant magnetic fields are observed in different regions of the circumstellar environment of several evolved stars, which indicate the possible role of the magnetic field in shaping the CSEs. Observations of H$_{2}$O masers revealed significant field strength for Mira variables and supergiants  in the range 0.2 G to 4 G \citep[][]{wouter2002,wouter2005}. In particular, H$_2$O maser polarimetric observations of the water fountain source W43A showed that the jet is magnetically collimated \citep[][]{wouternature}. Additionally, polarimetric observations of OH masers revealed large scale magnetic field  strength in evolved stars ranging from 0.1 mG to 10 mG \citep[e.g.][]{etoka2004,amiri2010}.

However, it should be noted that in interpreting the morphology of the maser emission in terms of the underlying physical structure a number of aspects have been ignored. This includes sufficient column density, velocity coherent path length along the amplification path and pumping mechanism. Therefore, it is possible that both SiO and OH molecules exist in spherical shells but the conditions for the masers mentioned above are not satisfied.

\subsection{SiO emission in OH~44.8-2.3}
The SiO masers of OH~44.8-2.3 exhibit a ring morphology (Fig. \ref{fig:OH44_ring}). Assuming a distance of 1.13$\pm$0.34 kpc, the masers are located at a distance of $\sim5.4$ AU from the surface of the star. The ring morphology is similar to the ring patterns observed previously for the SiO masers of Mira variables \citep[e.g.][]{diamond1994}. Moreover, the estimated SiO maser ring radius of OH~44.8-2.3 is similar to those measured for Mira variables \citep[e.g.3-7 AU;][]{cotton2008}. However, since OH~44.8-2.3 is a high mass loss OH/IR star, the CSE and stellar radius of this star is expected to be larger than Mira variables. The typical stellar radius for OH/IR stars corresponds to $\sim$2.8 AU \citep[][]{herman1985b}. This implies that the SiO masers of OH~44.8-2.3 occur at $\sim$1.9 stellar radius. This value is similar to the lower end location of SiO masers in Mira variables \citep[2-6 stellar radius;][]{elitzur1992}. Therefore, even though OH~44.8-2.3 is expected to be larger, the SiO masers occur at the same distance from the stellar photosphere as Mira variables.

However, the main uncertainty for the discussion above is the distance to the star. From the 1612 MHz OH maser observations of this source using the phase lag method a distance of 1.13$\pm$0.34 kpc was measured \citep[][]{langevelde1990}. This method assumes spherical symmetry of the OH maser shell. However, the distribution of the OH masers of this star (Fig. \ref{fig:OH1612_map}) does not indicate spherical expansion of the OH maser shell. Therefore, accurate determination of the distance of this source using the parallax method for the SiO maser features of OH~44.8-2.3 are necessary to understand the size and location of the SiO maser region.

\section{Conclusions}\label{conclusion ch6}
Our observations indicate a ring morphology for the SiO maser region of the OH/IR star OH~44.8-2.3. Assuming a distance of 1.13$\pm$0.34 kpc \citep[][]{langevelde1990}, the masers are located at $\sim5.4$ AU from the central star. The ring pattern is similar to that observed previously for Mira variables.

 The linear polarization morphology is consistent with the dipole magnetic field morphology in the SiO maser region of this star. However, we cannot rule out toroidal or solar type field morphologies. Polarimetric observations of the OH and H$_2$O maser regions of the CSE of this star are required to clarify the magnetic field morphology of this star.

 We report a tentative detection of circular polarization at $\sim$0.7$\%$ for the brightest SiO maser feature in the modest spectral resolution data set. However, due to the increased noise in high spectral resolution data set we can not confirm the detection. Further polarimetric VLBI observations of the SiO masers of this star with more integration time are necessary to clarify this. We note that based on polarization studies we can not distinguish between the Zeeman and non-Zeeman effects from the observations.

The SiO maser features of OH~44.8-2.3 exhibit two opposite arcs. Furthermore, the 1612 MHz OH masers of this star indicate an elongated shell morphology in the direction where there is a gap in the SiO maser emission. Additionally, the direction of the magnetic field is parallel or perpendicular to the location of the gaps in the SiO maser ring and the OH maser extent. This could be taken as a clue that there is a large scale magnetic field that imposes a preferred direction on the outflow over scales that span 2 orders of magnitude.  Because of the timescales involved in forming the OH shell, one would then conclude that the magnetic field is important imposing an asymmetric signature on the neutral outflow in the OH/IR phase. Furthermore, the high fractional linear polarization measured for the SiO masers of this star could indicate the possible role of the magnetic field in shaping the circumstellar environment of this star. However, we can not differentiate between the radiative or collisional pumping for the SiO maser ring of OH~44.8-2.3.

Follow up monitoring of the SiO maser region of OH~44.8-2.3 can give the distance through parallax measurements and this source offers a unique opportunity to check the phase-lag distance previously obtained by \cite{langevelde1990}. Additionally, the future observations will enable us to check the evolution of asymmetries in the CSE of this star.

\begin{acknowledgements}
This research program was supported by the ESTRELA fellowship, the EU Framework 6 Marie Curie Early Stage Training program under contract number MEST-CT-2005-19669. WV acknowledges support by the \emph{Deut\-sche
    For\-schungs\-ge\-mein\-schaft} through the Emmy Noether Research
  grant VL 61/3-1. We thank Bruce Partridge for giving us access to the EVLA observations of J2253+1608.
\end{acknowledgements}

\bibliographystyle{aa} 
\bibliography{reference.bib} 
\end{document}